\documentstyle[aps,prl,epsf,prb,twocolumn]{revtex}
\begin{document}
\twocolumn[\hsize\textwidth\columnwidth\hsize\csname @twocolumnfalse\endcsname

\draft
\title{Linear Scaling Solution of the Coulomb problem using wavelets}
\author{S. Goedecker}
\address{ Max-Planck Institute for Solid State Research, Stuttgart, Germany}
\author{O. V. Ivanov}
\address{ Max-Planck Institute for Solid State Research, Stuttgart, Germany}
\address{ and P.N. Lebedev Physical Institute, Moscow, Russia}
\date{\today}
\maketitle

\begin{abstract}      
The Coulomb problem for continuous charge distributions is a central problem in physics. 
Powerful methods, that scale 
linearly with system size and that allow us to use different resolutions in different 
regions of space are therefore highly desirable. Using wavelet based Multi Resolution 
Analysis we derive for the first time a method which has these properties. The power 
and accuracy of the method is illustrated by applying it to the calculation of 
of the electrostatic potential of a full three-dimensional all-electron Uranium dimer.
\end{abstract}
\pacs{PACS numbers: 02.60.Cb, 02.70.Rw, 31.15.-p}

]

\narrowtext
The theory of wavelets~\cite{meyer,daub} is one of the most important recent development in mathematics.
It allows one to apply a multi-scale analysis to problems that exhibit widely varying 
length scales~\cite{meyer}. Problems with this feature abound in all fields of physics. 
The problem we want to address here is the classical Coulomb 
problem for a continuous charge distribution $\rho$, i.e. we want to solve Poisson's equation 
\begin{equation} \label{poisson}
\nabla^2 V = -4 \pi \rho   \end{equation}
under the constraint that the potential $V$ vanishes at infinity. This basic equation can 
be found in nearly any field of physics and it is therefore essential to have 
efficient solution methods for it. There are two important 
requirements for an algorithm  that solves this problem. First it should scale linearly 
with the size of the charge distribution. Since in numerical applications the charge 
distribution is given on a grid a measure for the size of the charge distribution is 
the number of grid points necessary to represent it. This linear scaling property is 
of utmost importance because in many applications one needs grids consisting of a very large 
number of grid points. Second, it should allow for grids that are nonuniform, i.e. that have 
higher resolution in regions where this is required. 
For discrete charge distributions 
several algorithms~\cite{discrete} with these two properties exist and have become a standard for 
simulations  of large  coulombic and gravitational particle systems. For continuous 
charge distributions proposals have been put forward to map the continuous problem 
onto a discrete and use the above mentioned algorithms for discrete systems~\cite{chemistry}.
To the best of our knowledge, there exists however no linear scaling algorithm 
for nonuniform grids that can directly be applied to continuous charge distributions.
If one constrains oneself to uniform grids and periodic boundary conditions, there 
are of course the well known Fourier techniques~\cite{recipes}, that show a nearly 
linear $N log_2(N)$ scaling with respect to the number of grid points $N$. 
Non-periodic boundary conditions can be implemented in the context of Fourier techniques only 
by cutting off the long range Coulomb potential~\cite{eastwood}. 
Finite element methods allow nonuniform grids, but grid generation and preconditioning  
pose severe problems.
Using a basis of wavelet functions, we will present in this 
paper a method that scales strictly linear and allows for nonuniform grids.

There are many families of wavelets and one has to choose the most appropriate one 
for a specific application. A widely used family are the compactly supported orthogonal 
wavelets of Daubechies~\cite{daub}. The orthogonality property is convenient if one has to 
expand an arbitrary function in a basis of wavelets. Their disadvantage is that 
they are not very smooth, i.e. only a small number of derivatives is continuous.
One can however construct families of nonorthogonal wavelets that are much smoother.
In general the mapping from the numerical values on a grid to the expansion 
coefficients of the wavelet basis is rather complicated and slow for biorthogonal 
wavelets~\cite{daub}. 
An exception are the second generation interpolating wavelets~\cite{teter}, whose special 
properties allow us to do this mapping easily. In addition they give rise to 
a particularly fast wavelet transform~\cite{sgwlt}. 

Wavelets have already been successfully applied in several areas of physics~\cite{applicat}. 
In the context of electronic structure calculations 
a fairly small wavelet basis can describe the widely varying scales of both core and 
valence electrons~\cite{arias}. Self-consistent electronic structure calculations have also 
been done~\cite{wei}. In these self-consistent calculations the solution of Poisson's equation 
was however done by traditional Fourier techniques.

Let us now briefly review the theory behind biorthogonal wavelets~\cite{daub}. As in ordinary 
orthogonal wavelet theory there are two fundamental functions, the scaling function $\phi$ 
and the wavelet $\psi$. In the biorthogonal case there are however still the complementary 
scaling function $\tilde \phi$ and the complementary wavelet $\tilde \psi$.
Each scaling function and wavelet belongs to a hierarchical level of resolution.
By analysing a function with respect to these different levels of resolution one 
can do a so-called Multi Resolution Analysis (MRA).
The space belonging to a certain level of resolution $k$ is spanned by all the 
integer translations of the scaling function $\phi_{i,k}(x) \propto \phi( (\frac{1}{2})^k x-i)$. 
Any function can be expanded within this level of resolution.

\begin{equation} \label{expand}
f(x) \approx \sum_i s_{i,k} \phi_{i,k}(x) \; ,
\end{equation}

Since the scaling functions and their complementary counterparts are orthogonal 
$<\:\phi_{k,i}(x)\:|\:\tilde \phi_{k,j}(x)\:> = \delta_{ij}$,
the expansion coefficients $s_{i,k}$ are given by
\begin{equation} \label{excof}
s_{i,k} = <\: f(x)\:|\: \tilde \phi_{i,k}(x)\:>\:.
\end{equation}

The expansion~\ref{expand} becomes more accurate if one goes to a higher level of resolution,
i.e. if one decreases $k$ and it becomes exact in the limit $k \rightarrow -\infty$. 
This is an important feature because it allows us to 
improve systematically the numerical accuracy in very much the same way as it is done 
with a basis of plane waves. In numerical 
application it is of course not possible to take this limit, and we will therefore denote 
the finest level of resolution that is used in the calculation by $k=0$

The scaling function satisfies a refinement relation
\begin{equation}
 \phi_{j,k}(x) = \sum_l \tilde h_{l-2j} \phi_{l,k-1}(x) 
\end{equation}
i.e. each scaling function of a lower resolution level can be expressed as a linear 
combination of higher resolution scaling functions. It is obviously not possible 
to express a scaling function of higher resolution by a linear combination of 
lower resolution scaling functions only. One can, however, write down such an 
expression if one 
still includes the wavelets $\psi_{i,k}(x) \propto \psi( (\frac{1}{2})^k x-i)$

\begin{equation}  \label{trans2}
 \phi_{j,k-1}(x) = \sum_l h_{l-2j} \phi_{l,k}(x) + \sum_l g_{l-2j} \psi_{l,k}(x) 
\end{equation}

The wavelets at level $k$ thus reintroduces the resolution that is lost as one goes 
from level $k$ to level $k+1$ scaling functions.
These transformation properties among the scaling functions and wavelets 
give rise to the wavelet transform. The wavelet expansion coefficients $d_{i,k}$ 
are then either defined by this transform or equivalently by an expression 
analogous to Eq.~\ref{excof}.

\begin{equation}\label{forward}
s_{i,k+1}=\sum_j h_{i-2j} s_{j,k} \ ,\  d_{i,k+1} =\sum_j g_{i-2j} s_{j,k} 
\end{equation}
\begin{equation}\label{inverse}
s_{i,k} = \sum_j \left( \tilde h_{i-2j} s_{i,k+1} + \tilde g_{i-2j} d_{i,k+1} \right)
\end{equation}

Eq.~\ref{forward} is called a forward fast wavelet transform, Eq.~\ref{inverse} being 
its inverse counterpart. If one has periodic boundary conditions for the data $s_{i,k}$
the wavelet transform is one-to-one transformation between $s_{i,k}$ and 
its spectral decompositions $s_{i,k+1}$ and $d_{i,k+1}$. To obtain a full wavelet spectral 
analysis the forward transform is applied recursively. In consecutive steps 
the output data $s_{i,k+1}$  of the previous forward transform are the input data 
for the next transform. The size of the data set to be transformed is 
thus cut into half in each step. The total operation count is then given by a geometric 
series and scales therefore strictly linear. 
A full wavelet synthesis consists in the same 
way of a sequence of inverse transforms and gives back the original data set $s_{i,k}$. 
The coefficients $h_{i}$ and $g_{i}$ and their complementary counterparts 
$\tilde h_{i} \:, \tilde g_{i}$ are 
filters of finite length $2 m$ and can be derived from the MRA requirements~\cite{daub}.
The 8-th order lifted Lazy scaling function and wavelet~\cite{sgwlt} that were used in this 
work are shown in Fig.1. Because it can represent polynomials up to degree 8 exactly, 
the expansion coefficients with respect to the wavelets $d_{i,k}$ decay very rapidly for any 
smooth function with decreasing $k$.
    \begin{figure}             
     \begin{center}
      \setlength{\unitlength}{1cm}
       \begin{picture}( 5.0,3.0)           
        \put(-2.,-.45){\includegraphics{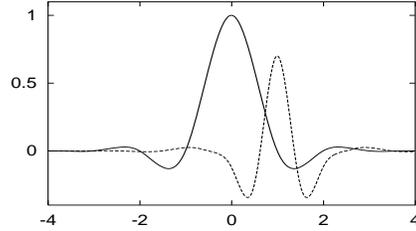}}   
       \end{picture}
       \caption{The scaling function (full line) and wavelet (dashed line) used in this work}
      \end{center}
     \end{figure}

To do a multidimensional MRA, we use a scheme described by Daubechies~\cite{daub}.
Even though all this work was done in the three-dimensional case, we will 
illustrate the principle just for the two-dimensional case.
The space of all scaling functions of resolution level $k$ is given by

\begin{equation} \label{ss}
\phi_{i,j,k}(x,y) = \phi_{i,k}(x) \phi_{j,k}(y) 
\end{equation}
The wavelet space is again defined as the space that recompensates for the resolution 
that is lost  by
going up one level in the scaling functions space. Using  Eq.~\ref{trans2} one obtains 
three kind of terms for the wavelet space
\begin{eqnarray}
 \psi^{01}_{i,j,k}(x,y) =  \phi_{i,k}(x) \psi_{j,k}(y) \label{sd}  \\
 \psi^{10}_{i,j,k}(x,y) =  \psi_{i,k}(x) \phi_{j,k}(y) \label{ds}  \\
 \psi^{11}_{i,j,k}(x,y) =  \psi_{i,k}(x) \psi_{j,k}(y) \label{dd}  
\end{eqnarray}

Let us now explain how to solve Poisson's equation in wavelets.
Expanding both the charge density and the potential in Eq~\ref{poisson}
into scaling functions 
\begin{eqnarray}
\rho(x,y,z) &=& \sum_{i,j} \rho_{i,j} \: \phi_{i,k}(x) \: \phi_{j,k}(y)  \\
V(x,y,z) &=& \sum_{i,j} V_{i,j} \: \phi_{i,k}(x) \: \phi_{j,k}(y)  
\end{eqnarray}
one obtains the following system of equations.
\begin{equation}
\sum_{j1,j2} L_{i,j;\mu,\nu} \: V_{\mu,\nu} = \rho_{i,j}
\end{equation}
where 
\begin{equation} \label{plin}
 L^k_{i,j;\mu,\nu} = 
<\: \tilde \phi_{i,k}(x)  \: \tilde \phi_{j,k}(y) 
              | \nabla^2 |  \phi_{\mu,k}(x)  \: \phi_{\nu,k}(y) \: >
\end{equation}
Since the scaling functions have a finite support, the matrix $L^k$ is a sparse matrix 
and its nonzero elements $L^k_{i1,i2;j1,j2}$ can be calculated analytically~\cite{beylkin}.
The natural boundary conditions for this scheme are periodic boundary conditions. 
As we stressed in the introduction, we however want to solve 
Poisson's equation~\ref{poisson} with non-periodic boundary conditions. As is well 
known, boundary affects vanish whenever the boundary is sufficiently far away. 
Thus, one could in principle obtain natural boundary conditions 
(i.e. $V(r) \rightarrow 0$ if $ r \rightarrow \infty $) within arbitrary precision 
if one uses a sufficiently large periodic box. Since the electrostatic potential 
decays fairly slowly a very large box is required and the numerical 
effort would be tremendous if one uses equally spaced grids within this huge periodic 
computational box. Far away from the charge distribution the variation of the potential is 
however small and less resolution is needed. The key idea is therefore to use 
a set of hierarchical grids as shown in Fig.2. where the resolution decreases as on goes out 
of the center. Expressed in the terms of wavelet theory this means that 
on the highest (periodic) level we have a basis of scaling functions. Resolution is 
then increased by adding wavelet basis functions near the center.  By doing this 
repeatedly we obtain increasing resolution towards the center as shown in Fig.2.

Up to now the motivation for introducing grids of different resolution was to 
handle the natural boundary conditions. Additional levels of resolution can however 
be introduced to handle charge distributions that have different length scales and require 
therefore higher resolution in some parts of space. The theory of wavelets gives us 
also enough flexibility to increase the resolution not only around one center but 
around any number of centers in the computational box.

    \begin{figure}             
     \begin{center}
      \setlength{\unitlength}{1cm}
       \begin{picture}( 5.0,3.5)           
        \put(-8.,-13.2){\includegraphics{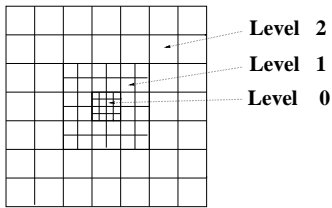}}   
       \end{picture}
       \caption{A hierarchical multi resolution grid of the type used in this work. 
                For simplicity only three levels of resolution are shown}
      \end{center}
     \end{figure}

A full wavelet synthesis step can be done straightforwardly in this hierarchical grid setting.
Any wavelet can be decomposed into scaling functions and therefore one can calculate 
the scaling function coefficients at any level of resolution and for any point in the 
computational volume. If one calculates these scaling functions for high resolution 
levels in a region of low resolution, one obtains however a highly redundant data set.
To do a full wavelet analysis that brings back the original spectral decomposition data, 
it turns out that one needs actually a slightly 
redundant data set. In order to calculate the wavelet coefficients for a wavelet 
at a boundary to a lower resolution region, one needs the scaling function 
values corresponding to this higher resolution also in a strip of width $m$ in the 
lower resolution region. A schematic diagram of a full hierarchical wavelet 
analysis and synthesis is shown in Fig.3. 

    \begin{figure}
     \begin{center}
      \setlength{\unitlength}{1cm}
       \begin{picture}( 5.0,7.5)           
        \put(-8.,-11.){\includegraphics{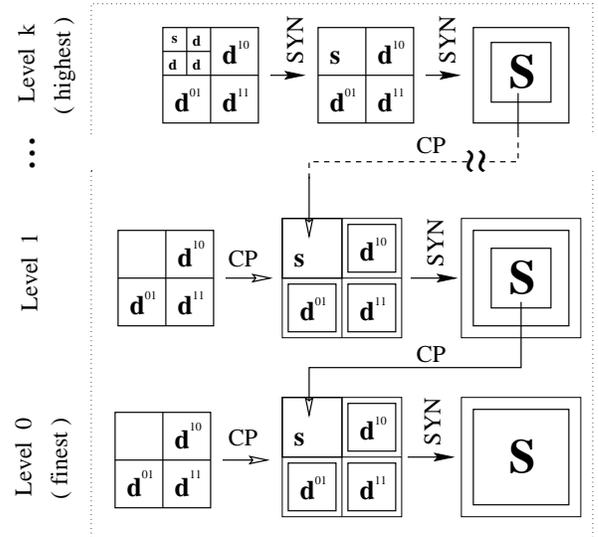}}   
       \end{picture}
       \caption{A schematic representation of a multi hierarchy wavelet synthesis.
                Data regions denoted by s,$d^{01}$,$d^{10}$ and $d^{11}$ contain 
                expansion coefficients for basis functions of the type given 
                by Eq.~\ref{ss}, \ref{sd}, \ref{ds} and \ref{dd} respectively.
                CP stand for a copy step where one puts an additional layer 
                of zeroes around the data set. SYN denotes a one level wavelet 
                synthesis step. One starts the process at the coarsest (periodic) 
                level and proceeds down to the finest resolution level. 
                To do a multi hierarchy wavelet analysis one proceeds back up
                reversing all the copy operations and replacing the single level 
                synthesis steps by analysis steps.  }
      \end{center}
     \end{figure}

In this mixed representation, where one has scaling functions at the highest periodic level 
and wavelets all the refinement levels, the structure of the Laplace operator is 
much more complicated since one has coupling between all the hierarchical levels.
An elegant way to cope with this additional complexity is the so-called 
nonstandard operator form proposed by Beylkin, Coifman and Rokhlin~\cite{nonstand},
which allows us to incorporate this coupling by a sequence of wavelet transforms (Fig.3), 
that are interleaved with the application of a simple one-level Laplace operator.
For this one-level Laplace operator only the matrix elements of
the Laplace operator among scaling functions and wavelets on the same
resolution level, but not between different levels of resolution are needed.

Mathematically the nonstandard operator form is a telescopic expansion of 
the Laplace operator in the scaling function basis at the finest level $L^0$.
If we define projection operators $P_k$ and $Q_k$, that project the whole space 
into the space of scaling functions and wavelets at the k-th level as well as 
their complementary counterparts $\tilde P_k$ and $\tilde Q_k$, they satisfy 
\begin{equation}
P_{k}= P_{k+1} + Q_{k +1} \hspace{.5cm} ; \hspace{.5cm} \tilde P_{k}= \tilde P_{k+1} + \tilde Q_{k +1}
\end{equation}
and we may write

\begin{eqnarray}
L^{k}= \tilde P_k \: L^0 \: P_k 
      & = & (\tilde Q_{k+1} + \tilde P_{k +1}) \: L^0 \: (Q_{k+1} + P_{k +1}) \nonumber \\
      & = & L^{k+1}_{DD} + L^{k+1}_{SD} + L^{k+1}_{DS} + L^{k+1}  \label{teles}
\end{eqnarray}
where $L^{k+1}_{DD}$, $L^{k+1}_{SD}$, $L^{k+1}_{DS}$ are Laplace operators at 
the $(k+1)$th level representing the coupling of wavelets with wavelets, 
wavelets with scaling functions and scaling functions with wavelets.
Applying Eq.~\ref{teles} recursively for $k=0,1,\ldots$, one obtains the nonstandard operator form.

In the basis of the wavelet functions at different resolution levels a simple
diagonal preconditioning scheme is very efficient and we were able to reduce the residue by
one order of magnitude with only 3 iterations.

To demonstrate the power of this method we applied it to a problem that can 
hardly be solved by any other methods, namely the potential arising from the nucleonic 
and electronic charge distribution of a fully three-dimensional all-electron 
Uranium dimer. The charge 
distribution of the nucleus was represented by a Gaussian charge distribution 
with an extension of $\frac{1}{2000}$ atomic units. Since the valence electrons 
have an extension which is of the order of one atomic unit, we have length scales 
that differ by more than 3 orders of magnitude. As can be seen from Fig.4, the 
potential also varies by many orders of magnitude. 
Using 22 hierarchical levels in our algorithm, we can 
represent resolutions that differ by 7 orders of magnitude and we are able to 
calculate the potential with at least 6 significant digits in the whole region 
from the nucleus to the valence region. 
In order to be able to determine the 
error we actually first fitted the electronic charge distribution by a small 
number of Gaussians, whose exact potential can be calculated analytically. 
This rather crude charge density was then used in all the calculations.

We also applied the method to clusters containing several $CO$ molecules that were 
described by pseudo-potentials. In this case there is only one length scale associated with 
the charge distribution and it is possible to reduce the number of grid points on 
higher levels such that the total amount of work increases only slightly with 
additional hierarchies. We were able to calculate the potential 
corresponding to the non-periodic boundary conditions with 8 significant digits.

We thank Mike Teter and Leonid Zaslavsky for bringing the beauty and 
usefulness of wavelets to our attention.
Jurg Hutter pointed out several essential references on 
wavelets to us. We acknowledge the interest of O. K. Andersen, O. Gunnarson, and M. Parrinello.

    \begin{figure}
     \begin{center}
      \setlength{\unitlength}{1cm}
       \begin{picture}( 5.0,5.0)           
        \put(-3,-.7){\includegraphics{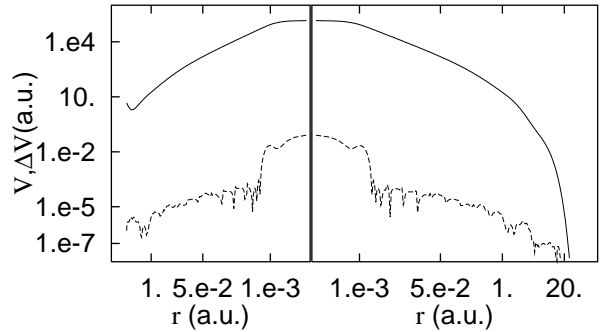}}   
       \end{picture}
       \caption{The potential $V$ (full line) and the numerical error $\Delta V$ 
                (dashed line) for an Uranium dimer  
                as a function of the distance from of right hand side nucleus. 
               The left panel shows both quantities in the direction of the left nucleus, 
	       the right panel shows them in the opposite direction. Both distances are 
               given on a logarithmic scale.}
      \end{center}
     \end{figure}

\end{document}